# Experimental evidence of giant chiral magnetic effect in type-II Weyl semimetal $WP_{2+\delta}$ crystals

Yang-Yang Lv[1†], Xiao Li[2†], Bin Pang[1], Y. B. Chen[2*], Shu-Hua Yao[1], Jian Zhou[1] & Yan-Feng Chen[1,3*]

[1] National Laboratory of Solid State Microstructures & Department of Materials Science and Engineering, Nanjing University, Nanjing 210093, China

[2] National Laboratory of Solid State Microstructures & Department of Physics, Nanjing University, Nanjing 210093, China

[3] Collaborative Innovation Center of Advanced Microstructures, Nanjing University, Nanjing 210093, China

[†] Y. Y. Lv and X. Li contribute equally to this work.

[*] Correspondence and requests for materials should be address to Y. B. Chen (ybchen@nju.edu.cn) or Y. F. Chen (yfchen@nju.edu.cn).

**Abstract**

Chiral magnetic effect is a quantum phenomenon that is breaking of chiral symmetry of relativistic Weyl fermions by quantum fluctuation under paralleled electric field *E* and magnetic field *B*. Intuitively, Weyl fermions with different chirality, under stimulus of paralleled *E* and *B*, will have different chemical potential that gives rise to an extra current, whose role likes a chiral battery in solids. However, up to now, the experimental evidence for chiral magnetic effect is the *negative longitudinal magnetoresistance* rather than a chiral electric source. Here, totally different from previous reports, we observed the giant chiral magnetic effect evidenced by: "negative" resistivity and corresponding voltage-current curves lying the second-fourth quadrant in type-II Weyl semimetal $WP_{2+\delta}$ under following conditions: the misaligned angle between *E* and *B* is smaller than 20°, temperature <30 K and externally applied *E*<50 mA. Phenomenologically, based on macroscopic Chern-Simon-Maxwell equation, the giant chiral magnetic effect observed in $WP_{2+\delta}$ is attributed to two-order higher coherent time of chiral Weyl-fermion quantum state over Drude transport relaxation-time. This work demonstrates the giant chiral-magnetic/chiral-battery effect in Weyl semimetals.

Chiral magnetic effect (CME) is a macroscopic quantum effect which is originated from the breaking of chiral symmetry in Weyl fermions by quantum fluctuation[1-5]. CME can be intuitively understood as follows: Weyl fermions are cataloged by their chirality-left or -right. Namely, their mechanical moment is paralleled/anti-paralleled to their spins in left-/right- chirality. Strikingly, the corresponding charge density of right-chirality is different from left-one if there are paralleled the *E* and *B*. In other words, the Weyl fermions with different chirality will have different chemical potential under the paralleled *E* and *B*, which will give rise to a current/flux[6,7]. This effect is also nominated as chiral battery effect in condensed matter[8,9]. The experimental observation of CME at particle-physics is focused on quark-gluon plasma under heavy-ion collision, which detects the CME in non-abelian gluon gauge field[5,8,10,11]. Dirac/Weyl semimetals, recently developed sub-division at condensed matter physics[12-18], provide a new table-top platform to explore CME in abelian electromagnetic gauge field[19-28]. Theoretically, whether there is CME in Weyl semimetals is under hot debate[25,28]. One of arguments is that the Hamilton in Weyl semimetal can be written to Weyl equation only around Weyl points within a limited energy range, rather than whole energy range at the high-energy particle physics[25,28]. Experimentally, only one work claims the observation of CME in $ZrTe_5$ through negative longitudinal magnetoresistance under paralleled *E* and $B$[19]. But it is a circumstantial proof of the CME because the generated current by CME perturbatively modifies the resistance (negative magnetoresistance) which is phenomenologically co-incident to Adler-Bell-Jackiw anomaly effect observed in TaAs, $Na_3Bi$, etc[29,30]. In short, the CME has not demonstrated chiral battery effect in Weyl semimetals.

Here we discovered the giant CME (chiral battery effect) evidenced by: "negative" resistivity and anomalous Ohm's law in a type-II 3D Weyl semimetal $WP_{2+\delta}$ crystals[31,32] (herein δ is dedicatedly controlled to be 0.11 to adjust their Fermi-level close its Weyl points). The *giant* CME here is to emphasize the significant CME effect in $WP_{2+\delta}$, to comparison, the magnetoconductivity increases in $ZrTe_5$ by ~400% over the zero-field value at 9 T and 5 K;[19] but, this value is ~40000% in $WP_{2+\delta}$ samples at the same condition described as follows. In these $WP_{2+\delta}$ single crystals, their nominal resistivity is negative and voltage-current curves are lying in the second-fourth quadrants instead of normally the first-third ones, under the following conditions: paralleled *B* and *E* fields, enough low temperature (< 40 K) and enough strong applied magnetic field (> 3 Tesla). Phenomenologically, this type of anomalous Ohm's law could be described by a $\theta$-field current density $\vec{j}_\theta$ generated by axion field in Maxwell-Chern-Simons electrodynamics[13,20,21].

**Results**

**Temperature-/magnetic-field-dependent anomalous electrical resistivity.** Fig. 1a is a temperature-dependent resistivity $\rho$ with various magnetic field *B* under *E* // *a* and *B* // *b*. Hereafter *a*, *b* and *c* are used to represent crystallographic directions of orthorhombic $WP_{2.11}$. Evidently, $WP_{2.11}$ is a typical metal without *B*. The residual resistance $\rho_{res}$ is 7.65 nΩ·cm at 2 K and the residual resistivity ratio (RRR) is calculated as about 2180, which suggests the extremely low defect concentrations. It is in agreement with reference that claims the largest mobility observed in $WP_2$ among the currently available compounds[33]. Under finite *B*, there is a metal-semiconductor-transition (MST) at $T_1$ ($T_1$~40 K), wherein, for $T<T_1$, $WP_{2.11}$ shows the semiconductor behavior (d$\rho$/d*T*<0). MST is commonly observed in Weyl semimetal $MoTe_2$/$WTe_2$[34], whose physical origin is still under hot debate. Inset of Fig. 1a is the optical micrograph of the electrodes used in electrical characterizations. Herein I- and II-points are source and sink of the external current; III- and IV-points are connected to a voltmeter. Fig. 1b shows the temperature-dependent $\rho$ with *E*//*a* and 9-Tesla *B* parallel to *a*, *b* and *c* of $WP_{2.11}$, respectively. In the $\rho$-*T* curve with *E* perpendicular to *B*, the MST happens at the low temperature. However, the most remarkable feature is that when *T*<40 K, resistivity evolves to negative under *E*//*B*. Conceptually, the resistivity $\rho$ represents the energy transferring from electrical carriers to other types of elementary excitations (phonon, magnon, etc.), or to crystal defects (point impurity, defects, etc.), which is always *positive* and determines the physical limit to the energy-consumption and carrier-mobility of electron-based devices resistivity[35]. How could it be possible that the resistivity in $WP_{2.11}$ evolves to negative under various temperature and *B*?

To address this wondering, we carried out the transport measurement further. Fig. 2 depicts the resistance *R-B* under different temperature *T* and the tilted *B* with *E // a*. It shows that when *E* is perpendicular to *B*, resistance difference $\Delta R=R(B)-R(0)$ is positive and quadratically dependent on *B* at various temperatures ( Fig. 2a & Fig. 2b), meanwhile its transverse magnetoresistance (MR) can reach as large as $1.65 \times 10^5$ % measured at 2 K with 9-Tesla B // *b*-axis. When *B // E* (Fig. 2c), $\Delta R$ remains positive when T>40 K, *but* evolves to negative when T<40 K with $\Delta R$ varying quadratically with the strength of magnetic field *B*. Quantitatively, the resistance of $WP_{2.11}$ are $3.35 \times 10^{-6} \Omega$ at B=0 and $-1.45 \times 10^{-3} \Omega$ at B=9 Tesla at T=2 K, respectively. Obviously, there is a giant negative *resistance* rather than negative *magnetoresistance*. The dependence of *R* on *B* with tilted *B* at 2 K is shown in Fig. 2d-2f. It shows that when *B* is tilted from *b* to *c*, *R* keeps positive and is parabolically dependent on *B*. Different from this case, when *B* is tilted from perpendicular to parallel to *E* (see Fig. 2e & 2f), *R* evolves from positive to negative when the mis-aligned angle between *B* and *E* is smaller than 20°. The parabolic relationship between *R* and *B* was observed as shown in Fig. 2e and 2f. These characterizations reveal definitely that the resistance of the $WP_{2.11}$ crystal indeed evolves from conventionally positive to negative under the low temperature, and mis-aligned angle <20°, and high applied magnetic field. It's too strange to be believable since negative resistance means no energy consumption and even getting gain in the crystals. What happens in it?

**Anomalous Ohm's Law.** In order to find the real things happened behind the observed negative resistivity, we conducted the direct *I-V* curve measurement. We measured *R-B* under different temperatures and tilted *B* under small steady-current of 15 mA. The current-voltage (*I-V*) curves (Fig. 3a-c) of the $WP_{2.11}$ crystal were obtained at 2 K with configuration of *E // a* under varying tilted angle φ of 9-Tesla *B*. When *E* is perpendicular to *B*, all *I-V* curves (see Fig. 3a) stay in the first-third quadrants, meanwhile the slopes of *I-V* curves are non-monotonously dependent on mis-aligned angle φ. It means that *R* is positive, the same as normal metals described by Ohm's law. Totally different from foregoing case, with 9-Tesla *B* tilted to *a*-axis, i.e. *B to E//a*, the slopes of *I-V* curves decrease dramatically with mis-aligned angle between *B* and *E* decreased, and at last the slopes of *I-V* curves (see Fig. 3b and 3c) evolve to negative. In other words, the fact corresponding to the negative resistivity is that the electric voltage is inversed from positive to negative. This observation is kept once the misaligned angle between *B* and *E* is less than 20°. Interestingly, the *I-V* curves in these measurements are located in the conventional first-third quadrants in accordance to Ohm's law when B ⊥E, but the *I-V* curves transform to second-fourth ones when *B* is nearly paralleled to *E*. The temperature dependent *I-V* curves, under *E* // 9-Tesla *B* // *a*, reveal that once T is lower than 40 K, the *I-V* curves could locate in abnormal second-fourth ones (Fig. 3d). Fig. 3e shows the *I-V* curves with varying *B* under *E//B//a*-axis at 2 K, which evidences that if external current is smaller than 15 mA, the *I-V* curves of WP$_{2.11}$ are linear. In the meantime, only when *B* is larger than 3 Tesla, are *I-V* curves located at the second-fourth quadrants. In other words, the negative resistivity could be maintained at enough strong *B* paralleled to *E*. Fitting these linear *I-V* curves gives us nominal negative resistance *R*. Plotting the extracted resistance *R* verse *B* in the inset of Fig. 3e, we get the parabolic relationship between *R* and *B*, which is in agreement to *R-B* curves in Fig. 2. The extracted parameters in Fig. 3e are almost the same as those in Fig. 2. It should be noticed that in the measurement of Fig. 3a-3e, the maximum of external current is 15 mA. What would happen to the *I-V* characters when applied large external current? Fig. 3f depicts the *I-V* curves of WP$_{2.11}$ with large external current *I* up to 80 mA, under varying strength of *B* with *B* // *E* //*a*. Let's trace the voltage evolution of *I-V* curve with *B* as 5-Tesla, the voltage decreases with current increased initially, then reaches a minimum when *I* is about 50 mA, and then increases with further increasing *I*, and at last, recovered to normal Ohm's law with $\frac{dV}{dI} > 0$. Other *I-V* curves show the similar trend. In order to analyze this evolution behavior, we fitted the *I-V* curves in Fig. 3f by a model of series connection of two resistors. The first resistor is normal including two terms through which the voltage is $a'\cdot I + b\cdot I^3$ (*a'* and *b* are two fitting parameters, and *I* is current). The *b* is attributable to the Joule heating effect which dominates *I-V* curve under large external current, which physical mechanism can be found in supplementary materials. The second resistor is an "active" component whose electrical behavior can be phenomenologically described as $-R_{anom}\cdot I$ ($R_{anom}$ is anomalous resistance). The fitting results are shown by line comparing with dots from experiments (see Fig. S5). These *I-V* characterizations confirm that the anomalous Ohm's law in $WP_{2.11}$ under *E // B*, could be found with the temperature lower than 40 K, the mis-aligned angle of *E* and *B* smaller than 20 degree, the magnetic field *B* greater than 3 Tesla, and external current smaller than 50 mA.

**Discussion**

The *possible* mechanism leading to the anomalous Ohm's law observed in $WP_{2.11}$ crystals might be augmented as following. Because all above-mentioned measurements are related to *macroscopic* parameters, we take the macroscopic electrodynamics as the starting point of discussion[11,20,21,24]. It has been proposed that the non-trivial Berry phase at the reciprocal space of Weyl semimetal results in the axion field, which could couple to macroscopic electromagnetic field described by the axion angle $\theta(\vec{r},t)$. In the Weyl semimetal, the axion angle $\theta(\vec{r},t)$ is equal to $\Delta\vec{k}\cdot\vec{r} - \Delta\omega\cdot t$, where $\Delta\vec{k}$ and $\Delta\omega$ are momentum and frequency separation between two Weyl nodes, respectively[11,20,21,24]. Our first-principles calculations (Fig. S6 & S7 at the supplementary information) verify that $WP_2$ is a type-II Weyl semimetal, which is the same as that reported in reference[32]. Therefore, it is reasonable to observe the effect of axion field in $WP_{2.11}$ crystals wherein doping $WP_2$ with more P as high as 0.11 mole can move the Fermi level of $WP_{2.11}$ close to the corresponding Weyl points. Taken into account of the axion angle $\theta$-term, the Lagrangian *L* can be written as

$$L = L_{\text{Maxwell}} + \frac{e^2}{4\pi^2} \cdot \theta(\vec{r},t) \vec{E} \cdot \vec{B} \qquad (1)$$

in Maxwell-Chern-Simons electrodynamics, where $L_{\text{Maxwell}}$ represents the normal Maxwell Lagrangian, $e$ is the electron charge. Here, $\theta$-term modifies the two of conventional Maxwell equations. The Maxwell-Chern-Simons equation relevant to our experiment is:

$$\nabla \times \vec{B} = \frac{4\pi}{c} \vec{j}_{external} + \frac{1}{c} \frac{\partial \vec{D}}{\partial t} - \frac{e^2}{2\pi^2} \frac{\partial \theta}{\partial t} \vec{B} \qquad (2)$$

where the time-derivation of $\vec{D}$ is the displacement current; $\vec{j}_{external}$ is external current density. This formula suggests that due to the existence of axion field, a $\theta$-field current density $\vec{j}_\theta$ ( $\vec{j}_\theta = -\frac{e^2}{2\pi^2 \hbar^2} \Delta\varepsilon \vec{B}$, $\hbar$ is the Plank constant) will be generated under paralleled $E$ and $B$. $\Delta\varepsilon$ is the energy difference between positive- and negative-chirality electrons ($\Delta\varepsilon = \hbar \cdot \Delta\omega$)[11,19]. Considered chiral carrier concentration difference generated by paralleled $E$ and $B$[11,19], $\Delta\varepsilon$ can be written as $\frac{3\hbar}{4\pi^2}(v \cdot \tau_c) \frac{v^2 \vec{E} \cdot \vec{B}}{\mu^2}$, (where $v$, $\tau_c$ and $\mu$ are Fermi velocity, *de-coherence time* and Fermi energy, respectively). It's interesting to notice that the sign of $\Delta\varepsilon$ depends on $E$ parallel or anti-parallel to $B$. It could qualitatively explain why the *I-V* curves locate in VI or II quadrant when $B$ is parallel or anti-parallel to $E$. Then the θ-field current density $\vec{j}_\theta$ for paralleled E and B is:

$$\vec{j}_\theta = -\frac{e^2}{8\pi^4 \hbar}(v \cdot \tau_c) \left(\frac{evB}{\mu}\right)^2 E \qquad (3)$$

It should be emphasized that this formula is quite similar to that in Adler-Bell-Jackiw (ABJ) anomaly[7], but the physical content is completely different. For ABJ anomaly, the effect of $\theta$-field current can be renormalized to the resistivity, however, the $\theta$-field current in our experiment provides a *current source*. The key point to distinguish these two cases is based on the physical meanings of $\tau_c$: it is explained as a *relaxation time* of scattering between Dirac cones with different chirality in ABJ anomaly[7], while in this work, $\tau_c$ should be understood as *de-coherence time of the Weyl fermion quantum state*[36]: the average duration when positive-/negative- chirality electrons could keep their phase memorized. In other words, $\tau_c$ here is the averaged time over which the electrons pumped to high energy states by $E \cdot B$ will forget their chirality and fall back to their original states. As Eq. (3) suggests, greater $\tau_c$ leads to larger current density $\vec{j}_\theta$. Phenomenologically, the current $\vec{j}_\theta$ generated by large $B//E$ at low temperature will surpass the externally applied current. The externally applied current, in the measurement, is provided by a constant-current source. To keep the constant-current assigned to the whole circuit, the constant-current source has to apply "negative" voltage upon WP$_{2.11}$ crystals to restrain the $\vec{j}_\theta$, which leads to the measured "negative" resistivity and anomalous Ohm's law. Based on above discussion, the physical parameters we really measured are electrical behaviors of an "active" electrical device (like a battery, chiral battery) under either fixed external electric-field $E$ or fixed magnetic-field $B$. Two typical curves shown at Fig. 2 and 3 are re-plotted in the Fig. 4. Evidently, $\vec{j}_\theta$ is either proportional to $B^2$ under the fixed $E$ or proportional to $E$ under the fixed $B$, which is in agreement to the equation (3). Fitted the curve in Fig. 4a by equation (3), the order of magnitude of *de-coherence time* $\tau_c$ at 2 K can be estimated as $10^{-9}$ s (the $\mu$ and $v$ are chosen as 1.0 meV and $1.0 \times 10^6$ m/s in the calculation, respectively). At the same time, the conventional relaxation time $\tau_0$ of WP$_{2.11}$ is about $10^{-12}$ s based on Drude model ($\tau_0 = \frac{m\sigma}{ne^2}$, here σ is the conductivity; $n$ and $m$ are carrier concentration and mass of free-electron, respectively[35]). Thus, the de-coherence time $\tau_c$ is two-order larger than $\tau_0$, which infers that positive-/negative- chirality electrons have long enough time to build large enough $\Delta\varepsilon$ to generate significant $\vec{j}_\theta$. With the temperature increased, the *de-coherence time* $\tau_c$ is decreased and close to zero at 50 K (see Fig. S4). The temperature-sensitive $\tau_c$ may be attributed to quantum coherence loss under thermal perturbation, as well as Fermi level deviated far away from Weyl points under thermal agitation. These discussions suggest that main features of our experimental results can be described by $\vec{j}_\theta$ generated by axion field in Weyl semimetals, and its effect is more like an electrical source (chiral battery effect)[8,9], rather than a modified resistivity as those reported in $Na_3Bi$[29]. Obviously, above-discussions are qualitative/semi-quantitative. We believe that above-reported experimental data could excite the further theoretical/experimental work to explore/elucidate the extraordinary effect of axion field in the condensed matter physics systems.

**Methods**

**The growth, measurement and characterization.** Details of $WP_{2\pm\delta}$ crystals can be found in the supplementary information. It should be emphasized that there are Weyl points in $WP_2$ lying around -260 meV[32] below Fermi level. Around the Weyl points there are linear dispersions between energy and momentum. Accordingly, there is finite Berry curvature that is the sufficient and necessary condition to have finite axion field[21-25]. To downshift to this Fermi level, we dedicatedly adjusted growth parameters to obtain $WP_{2+\delta}$ crystals with $\delta \approx 0.11$ which is determined experimentally as described in the supplementary information. These crystals demonstrate the exotic electrical properties described as follows. As shown in Fig. S1, X-ray diffraction (XRD), transmission electron microscopy (TEM) and elemental mapping characterizations prove that $WP_{2.11}$ samples have single-crystalline quality and uniform elemental distribution. The samples characterized by XRD and TEM are the same as those used in electrical measurement.

**The transport measurements.** Standard four-probe method was employed on the rectangular crystals for the electrical transport measurements which were performed in a 9 Tesla physical properties measurement system (PPMS-9T, Quantum Design). During the electrical measurement, the electrical current is aligned along [100]-direction and the electrodes were covered the whole samples to avoid the current-injecting effect[37,38]. The *I*–*V* curves were investigated using 2612B system SourceMeter (Keithley Inc.). The electrical current used in the measurement is DC.

The presented data are original; we have not averaged values at plus/minus current or plus/minus magnetic field *B*.

28. Hosur, P. & Qi, X. L. Recent developments in transport phenomena in Weyl semimetals. *Comptes Rendus Physique* **14**, 857 (2013).

29. Xiong, J. et al. Evidence for the chiral anomaly in the Dirac semimetal $Na_3Bi$. *Science* **350**, 413-416 (2015).

30. Huang, X. et al. Observation of the Chiral-Anomaly-Induced Negative Magnetoresistance in 3D Weyl Semimetal TaAs. *Phys. Rev. X* **5**, 031023 (2015).

31. Soluyanov, A. A., Gresch, D., Wang, Z. J., Wu, Q. S., Troyer, M., Dai, X. & Bernevig, B. A. Type-II Weyl semimetals. *Nature* **527**, 495-498 (2015).

32. Autes, G., Gresch, D., Troyer, M., Soluyanov, A. A. & Yazyev, O. V. Robust Type-II Weyl Semimetal Phase in Transition Metal Diphosphides $XP_2$ (*X*=Mo, W). *Phys. Rev. Lett.* **117**, 066402 (2016).

33. Kumar, N. et al. Extremely high magnetoresistance and conductivity in the type-II Weyl semimetals $WP_2$ and $MoP_2$. *Nat. Commun.* **8**, 1642 (2017).

34. Lv, Y. Y. et al. Experimental Observation of Anisotropic Adler-Bell-Jackiw Anomaly in type-II Weyl Semimetal $WTe_{1.98}$ Crystals at the Quasiclassical Regime. *Phys. Rev. Lett.* **118**, 096603 (2017).

35. Ziman, J. M. *Electrons and Phonons: The Theory of Transport Phenomena in Solids*. (Oxford University Press, 1958).

36. Annett, J. F. *Superconductivity, Superfluids and Condenstates*. (Oxford University Press, 2004).

37. Pippard, A. B. *Magnetoresistance in Metals*. (Cambridge University Press, 1989).

38. Zhang, C. -L. et al. Signatures of the Adler-Bell-Jackiw chiral anomaly in a Weyl fermion semimetal. *Nat. Commun.* **7**, 10735 (2016).13

**Acknowledgements**

We should acknowledge the financial support from National Key Basic Research Program of China (2015CB921203, 2013CB632702), and the Nation Science Foundation of China (11374149, 51032003, 50632030, 51002074, and 51472112). We'd like to thank Prof. X. G. Wan at Nanjing University for his enlightening discussions, Dr. X. J. Yan's experimental guidance for the *I-V* curves, and Prof. X. P. Liu's read and revises to our draft. Y. -Y. Lv acknowledges the financial support from the program A for Outstanding PhD candidate of Nanjing University (201701A008).

**Author contributions**

Y.-Y.L. and S.-H.Y. performed the crystal growth in assist of B.P.. Y.-Y.L. and Y.-B.C. determined the structure content and the elemental composition of the crystals. Y.-Y.L. and X.L. conducted the transport measurements. Y.-Y.L., X.L. and Y.-B.C. analyzed the data and refined the measurements. J.Z. and Y.-F.C. contributed to the result analysis. Y.-B.C. and Y.-Y.L. co-wrote the manuscript. S.-H.Y., J.Z. and Y.-F.C. revised the manuscript. All authors discussed the results and commented on the manuscript.

**Additional information**

**Competing financial interests:** The authors declare no competing financial interests.

**Figures captions**

**Fig. 1** The temperature dependent resistivity $\rho$ of WP$_{2.11}$ under magnetic field $B$ aligned along $a$, $b$ and $c$-axis of WP$_{2.11}$. **a** $\rho$-T relationship with $B$ varying from 0 to 9 Tesla with $E // a$ and $B // b$. One can see that there is metal-semiconductor-transition at low temperature T (T<$T_1$ ~ 40 K). Inset is the optical micrograph showing the electrodes used in the electric measurement. I- and II- are source and sink of external current, III- and IV-electrode were connected to a voltmeter. **b** $\rho$-T curves under $E // a$, with 9-Tesla $B // a$, $B // b$ and $B // c$, respectively. The inset is the measurement configuration. The negative resistivity is observed when T<40 K under 9-Tesla $B // E // a$.

**Fig. 2** The relationship between resistance $R$ of WP$_{2.11}$ and $B$ with varying temperatures under tilted and non-tilted $B$ with $E//a$. **a** The resistance change $\Delta R$-$B$ curves under varying T, with $B // b$, and (**b**) with $B // c$, and (**c**) with $B // a$, respectively. The negative $\Delta R$ is observed when $E // B // a$. **d** The $R$-$B$ curves with $B$ tilted from $b$ to $c$ axis at 2 K. **e** with $B$ tilted from $b$ to $a$ axis. **f** with $B$ tilted from $c$ to $a$ axis. All insets are the measurement configurations. The negative resistance is observed when mis-aligned angle between $B$ and $E$ is smaller than 20°, and temperature <40 K.

**Fig. 3** $I$-$V$ curves of WP$_{2.11}$ under tilted $B$, varying temperature, and different external current. **a** $I$-$V$ curves of WP$_{2.11}$ with $B$ tilted from $b$- to $c$-axis. **b** with $B$ tilted from $b$- to $a$-axis. And **c** with $B$ tilted from $c$- to $a$-axis. **d** $I$-$V$ curves of WP$_{2.11}$ under different temperatures, with 9-Tesla $B // E // a$. All insets are measurement configurations. Obviously, $I$-$V$ curves are located at the second-fourth quadrants when T<40 K and mis-aligned angle between $B$ and $E$ is less than 20°. **e** The linear dependence between $V$ and $I$ under various B and $I$ < 15 mA. Inset is $B$-dependent resistance $R$, extracted from linear-fitting of curves in Fig. 3e. $R$ is quadratically dependent on $B$, which is the same as those shown in Fig. 2. **f** $I$-$V$ curves of WP$_{2.11}$ under $I \leq 80$ mA, $B$ increased from 0 to 5 Tesla at 2 K, and $B // E // a$. Evidently, the $I$-$V$ curves locate at the second-fourth quadrants when B $\geq$ 3-Tesla. The conventional positive resistivity

(positive slope of *I-V* curve) is recovered when the external current is larger than 45 mA.

**Fig. 4** The abnormal Ohm's law observed in WP$_{2.11}$ could be understood as a current $j_\theta$ generated by the axion-field in Weyl semimetals under paralleled *E* and *B*. **a** Re-plot $\rho$-*B* curve (*B* // *E* // *a* and T=2 K shown in Fig. 2c) as $\frac{(j_\theta + j_{external})R_0 S}{I_{external}}$ -*B* relationship. Where $j_{external}$ is externally applied current density, $R_0$ and *S* are normal resistance of WP$_{2.11}$ and cross-section area of sample, respectively. This curve shows that $j_\theta$ is proportional to $B^2$ under fixed external electric-field *E*. **b** Re-plot *I-V* curve (9-Tesla *B* // *E* // *a* and T=2 K shown in Fig. 3e) as $(j_\theta + j_{external})R_0 S$ - $I_{external}$ relationship. It reveals that $j_\theta$ is negatively proportional to *E* under fixed applied magnetic-field *B*. These features are in agreement to equation (3) at the main text.

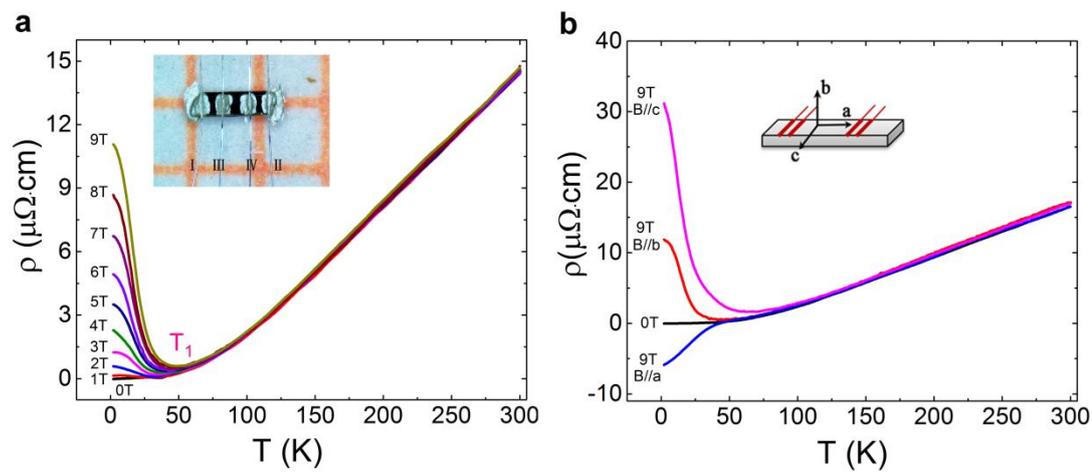

**Fig. 1**

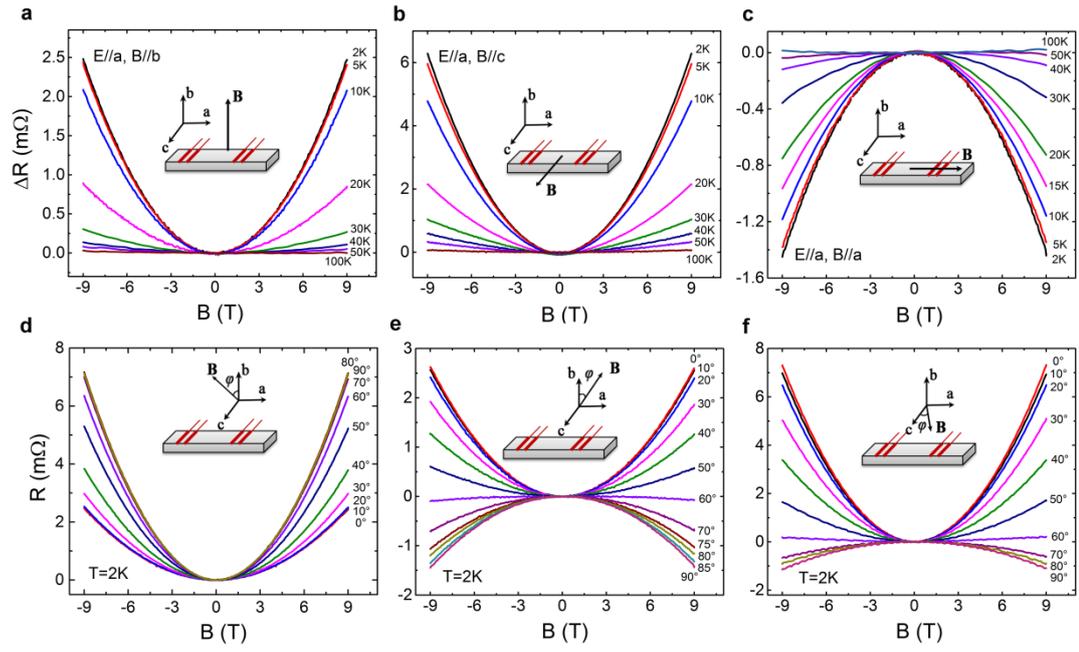

**Fig. 2**

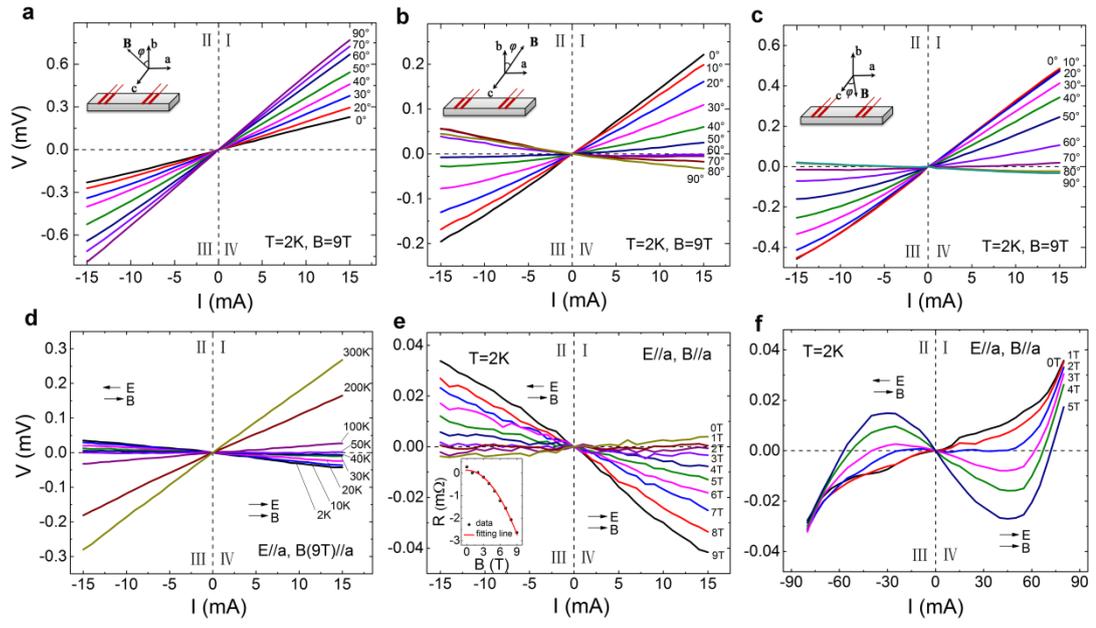

**Fig. 3**

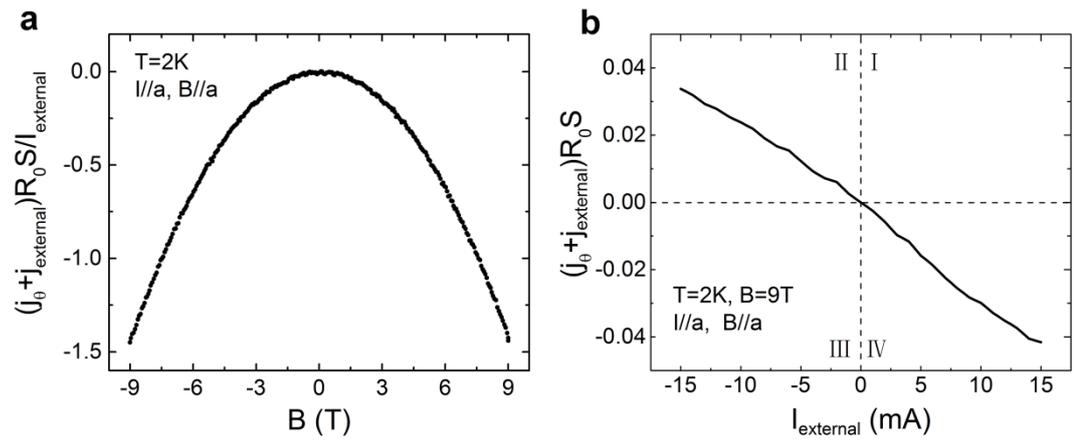

**Fig. 4**